\documentclass[journal]{IEEEtran}
\IEEEoverridecommandlockouts

\usepackage[utf8]{inputenc} 
\usepackage{times}

\usepackage{mathtools} 
\usepackage{amssymb}   
\usepackage{mathrsfs,bm,cases}
\usepackage{pifont,bbding,wasysym,fontawesome}

\usepackage[table]{xcolor}
\usepackage{array,booktabs,tabularx}
\usepackage{longtable,multirow,threeparttable,makecell}

\usepackage{graphicx,epsfig,psfrag}
\usepackage{subfigure} 
\usepackage{float,cuted,stfloats,balance}

\usepackage{algorithm,algorithmic}
\usepackage[algo2e,ruled,vlined,linesnumbered,commentsnumbered]{algorithm2e}

\usepackage{enumerate,verbatim}
\usepackage{cite,url}

\usepackage[
  colorlinks=true,
  linkcolor=.,
  citecolor=.,
  urlcolor=.
]{hyperref}

\begin{document}

\title{AFDM-Enabled ISAC in Dynamic Environments: Fundamentals, Technologies and Opportunities}

\author{Linchu Chen, Zhendong Li, Zhou Su, Lin Chen, and Wen Chen
\thanks{Linchu Chen and Zhendong Li are with the School of Information and Communication Engineering, Xi'an Jiaotong University, Xi'an 710049, China (e-mail: chenlinchu@stu.xjtu.edu.cn; lizhendong@xjtu.edu.cn). Zhou Su is with the School of Cyber Science and Engineering, Xi'an Jiaotong University, Xi'an 710049, China (e-mail: zhousu@ieee.org). Lin Chen is with the Department of Electrical and Computer Engineering, Stevens Institute of Technology, Hoboken, NJ 07030, USA (e-mail: lchen53@stevens.edu). Wen Chen is with the School of Integrated Circuits, Shanghai Jiao Tong University, Shanghai 200240, China (e-mail: wenchen@sjtu.edu.cn). (Corresponding author: Zhendong Li).}
\vspace{-1.5em}}

\maketitle

\begin{abstract}
Dynamic environments pose fundamental challenges to integrated sensing and communication (ISAC), particularly due to severe Doppler effects, and rapidly time-varying channels. Affine frequency-division multiplexing (AFDM), with its inherent capability of characterizing and separating delay and Doppler effects, has emerged as a promising waveform. This article provides a comprehensive overview  {of} AFDM-enabled ISAC in dynamic environments, covering its fundamental principles, distinctive advantages, representative application scenarios, and key enabling technologies. We first characterize the key features of ISAC in dynamic environments and introduce the fundamentals of AFDM, followed by an analysis of scenarios where AFDM can provide significant performance benefits.  {We then discuss key enabling technologies for AFDM-based ISAC in dynamic environments and examine parameter estimation and predictive beam tracking in a vehicular case study.} Finally, open challenges and promising future research directions are discussed, aiming to provide a comprehensive reference for researchers and practitioners while inspiring further innovation in this emerging field.

\end{abstract}

\section{Introduction}
\IEEEPARstart{I}{ntegrated}  { sensing and communication (ISAC) is a key capability for future wireless systems, especially when communication links and sensing targets move.} Relative motion among transceivers, targets, and scatterers produces time-varying path delays and Doppler shifts.  {When these variations are significant, the doubly dispersive channel challenges both data detection and timely sensing. Orthogonal frequency-division multiplexing (OFDM) suffers Doppler-induced inter-carrier interference (ICI) as the channel varies within a block~\cite{afdmcore}. These challenges motivate waveforms that help the receiver distinguish propagation paths and exploit them.}

 {As a chirp-based multicarrier waveform, affine frequency-division multiplexing (AFDM) builds on the discrete affine Fourier transform (DAFT). Its configurable chirp parameters determine how path delays and Doppler shifts appear in the DAFT-domain effective channel.} With parameters matched to the channel, paths with different delay-Doppler pairs can occupy distinguishable offsets, supporting structured channel estimation and detection~\cite{afdmcore}.  {Under suitable channel and detection conditions, AFDM can achieve full path diversity. The same path responses also carry delay and Doppler information for sensing. The receiver can estimate these parameters from an embedded DAFT-domain pilot and its guard region within a data frame~\cite{afdmsense}. Chirp design therefore governs both the paths available for data detection and the resources needed for sensing. This coupling between communication and sensing underpins AFDM's robustness to mobility.}

Beyond its robustness to mobility, the chirp-based structure of AFDM naturally lends itself to integrated communication and sensing. In particular, its radar-like ambiguity function determines range and velocity estimation and enables accurate, unambiguous sensing. Existing work \cite{RadarCentric} jointly realized communication and sensing by embedding information in the chirp parameters while maintaining favorable sensing performance. This flexibility stems from the additional design degrees of freedom in the DAFT domain, making AFDM promising for ISAC. These parameters can also enhance physical-layer security. In \cite{ChPer}, a time-varying chirp parameter-hopping AFDM waveform was investigated for secure satellite-air integrated communications. AFDM thus offers a unified design space for communication, sensing, and security.

 {These properties make AFDM a promising candidate for dynamic ISAC in practical scenarios. The four settings differ in their motion patterns, propagation geometry, and receiver constraints, and AFDM exhibits tailored advantages in response to these specific scenario characteristics. For instance, existing studies have investigated AFDM-based preambles for random access under satellite link frequency offsets~\cite{LEO} and exploited structured channel responses for low-complexity estimation in vehicular networks~\cite{VTX1}. Despite these advances, research on deploying AFDM-enabled ISAC across dynamic scenarios remains at an early stage. Existing overviews have examined how AFDM adapts to different propagation conditions~\cite{yin2025} and analyzed its communication diversity, ambiguity functions, estimation bounds, and sensing algorithms~\cite{yin2026}.}  {Nevertheless, a comprehensive analysis linking these theoretical aspects to specific application scenarios and key enabling technologies deserves further exploration.}

The remainder of this article is organized as follows.  {First, the prominent features of ISAC in dynamic environments are characterized, and several representative waveforms are introduced, with particular emphasis on AFDM.}  {Subsequently, several typical dynamic scenarios are presented and their characteristics analyzed, with the advantages of AFDM highlighted.}  {Furthermore, several key enabling technologies for AFDM-enabled ISAC are examined, and a case study is conducted in a typical vehicular network scenario.}  {Finally, open challenges and future directions are discussed.}

\section{ {Waveform Fundamentals and Selection for ISAC in Dynamic Environments}}

 {This section examines waveform requirements for dynamic ISAC and compares representative candidates.}  {We first analyze propagation conditions that distinguish dynamic from quasi-static ISAC and identify OFDM's limitations under severe Doppler.}  {We then introduce OFDM, orthogonal time frequency space (OTFS), orthogonal chirp division multiplexing (OCDM), and AFDM, and discuss path separation, complexity, and suitability.}

\subsection{ {Waveform Design Requirements in Dynamic ISAC}}
The evolution of ISAC toward increasingly dynamic application scenarios has created new challenges for both communication and sensing.  {Relative motion among transceivers, targets, and scatterers makes propagation vary within a block and between sensing updates.}  {These effects are pronounced in high-mobility environments such as vehicular networks.}  {Within a block, different propagation paths can exhibit different Doppler shifts even after compensation for a common frequency offset.}  {When residual Doppler spread is appreciable relative to the inverse block duration, the constant-channel assumption becomes inadequate.}  {Waveform choice then affects how readily the receiver distinguishes paths with different delays and Doppler shifts and exploits them for detection.}  {Between updates, changes in range, radial velocity, and bearing can make earlier estimates less useful for beam steering and resource allocation.}  {A shared ISAC waveform must therefore support timely sensing while carrying data.}  {These requirements motivate comparing candidate waveforms by path separability, sensing support, and pilot resources.}

\subsection{ {Technical Principles of Representative Waveforms}}
\label{sec:principles}

\begin{figure}
    \centering
    \includegraphics[width=0.8\linewidth]{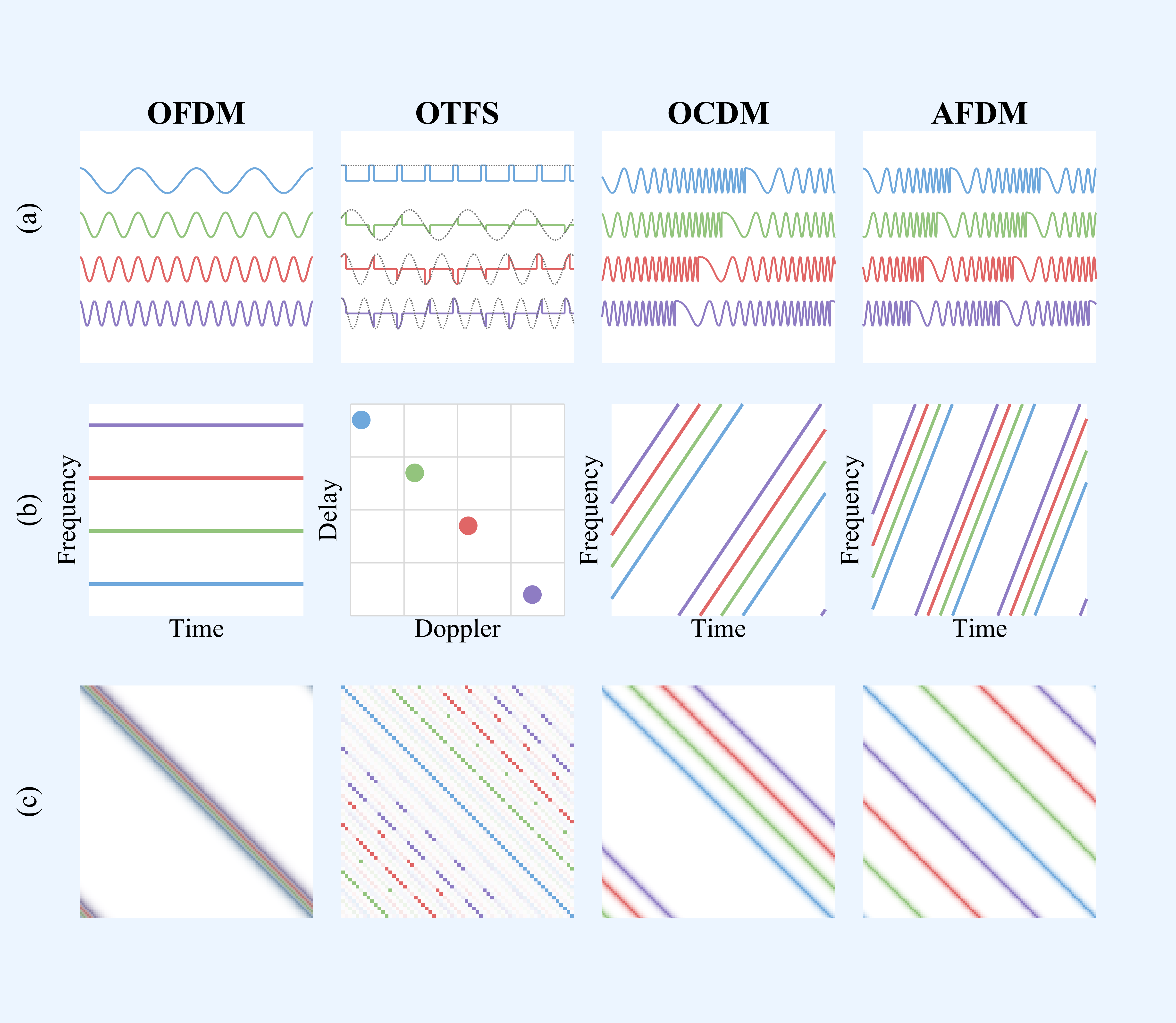}
    \caption{ {Comparison of the fundamental signal representations of four representative waveforms: (a) Time-domain basis; (b) TF occupancy; (c) effective channel.}}
    \label{yuanlitu}
\end{figure}
\subsubsection{ {OFDM}}
 {OFDM maps symbols onto orthogonal subcarriers, each with a complex-exponential basis.}  {For a block of $N$ samples, the $k$-th basis function is $\phi_k[n]=N^{-1/2}e^{j2\pi kn/N}$, where $k,n=0,\ldots,N-1$.}  {At the transmitter, an inverse fast Fourier transform (IFFT) combines the subcarriers into a time-domain block, and a cyclic prefix (CP) is appended to accommodate multipath delays.}  {If the CP covers the delay spread and the channel is roughly constant over the block, multipath becomes a circular convolution.}  {The receiver removes the CP, applies an FFT, and obtains a diagonal channel, so a single-tap equalizer recovers each symbol.}  {This efficient handling of frequency selectivity has made OFDM widely adopted.}  {In high-mobility scenarios, however, substantial Doppler spread causes the channel to vary within a block and couples the subcarriers, giving rise to ICI.}  {Beyond this impairment, OFDM's frequency-domain representation decomposes a frequency-selective channel into subcarriers but does not directly resolve paths in delay $\tau$ and Doppler $\nu$.}  {With conventional per-subcarrier detection, the receiver cannot fully exploit the diversity of all resolvable paths.}

\subsubsection{ {OTFS}}
 {OTFS places information symbols $x[k,l]$ on a two-dimensional delay-Doppler (DD) grid \cite{otfsModel}, where $k=0,\ldots,K-1$ and $l=0,\ldots,L-1$ denote the Doppler and delay indices, respectively.}  {The symbols are mapped to the time-frequency domain through an inverse symplectic finite Fourier transform, $X[n,m]=\frac{1}{\sqrt{NM}}\sum_{k=0}^{N-1}\sum_{l=0}^{M-1}x[k,l]e^{j2\pi(nk/N-ml/M)}$, and then converted into a time-domain signal for transmission.}  {At the receiver, time-frequency demodulation followed by a symplectic finite Fourier transform produces DD-domain observations.}  {Each path thus contributes a shifted, phase-rotated copy of the DD grid  in Fig. \ref{yuanlitu}(b).}  {When few paths have stable delays and Doppler shifts over a frame, the receiver can exploit this structure for joint detection.}  {This makes OTFS attractive for fast time-varying channels.}  {However, joint detection over the two-dimensional delay-Doppler grid incurs high computational complexity.}  {Deployment therefore requires balancing channel dynamics, estimation tolerance, complexity, and latency.}

\subsubsection{ {OCDM}}
 {OCDM places information symbols in the discrete Fresnel domain and maps them onto a set of orthogonal chirp subcarriers \cite{ocdmmodel}.}  {For an even block length $N$, the $k$-th chirp basis function can be written as $\psi_k[n]=N^{-1/2}e^{j\pi/4}e^{-j\pi(n-k)^2/N}$, where $k,n=0,\ldots,N-1$.}  {At the transmitter, the inverse discrete Fresnel transform (IDFnT) combines the chirps into a block and adds a cyclic prefix for multipath delays.}  {The IDFnT can be implemented using Fourier transforms and phase rotations.}  {The receiver removes the prefix and applies the discrete Fresnel transform to demodulate.}  {Each chirp spans the full bandwidth, so a symbol is distributed across frequencies and can exploit frequency-selective fading through equalization.}  {Furthermore, OCDM exhibits a certain capability for path separation in the Fresnel domain.}  {However, in doubly selective channels, path contributions may overlap in the Fresnel domain.}  {Because its chirp structure is fixed, path overlap degrades performance.}

\subsubsection{ {AFDM}}
 {AFDM places information symbols in the DAFT domain and maps them onto a set of orthogonal chirp subcarriers.}  {For a block of $N$ samples, the time-domain signal is given by $s[n]=N^{-1/2}\sum_{m=0}^{N-1}x[m]e^{j2\pi(c_1n^2+mn/N+c_2m^2)}$, where $x[m]$ is the $m$-th DAFT-domain symbol and $c_1$ and $c_2$ are the chirp parameters.}  {The inverse DAFT uses an IFFT with phase rotations before and after, followed by a chirp-periodic prefix for multipath delays.}  {The receiver removes the prefix and applies the DAFT, yielding $\mathbf{y}=\mathbf{H}_{\mathrm{eff}}\mathbf{x}+\mathbf{w}$, where $\mathbf{H}_{\mathrm{eff}}$ is the DAFT-domain effective channel and $\mathbf{w}$ is the noise.}  {The AFDM waveform maps each path's delay-Doppler pair to an offset along a diagonal of $H_{\mathrm{eff}}$ (Fig. \ref{yuanlitu}(c)).}  {Selecting $c_1$ for the delay and Doppler ranges maps the $r$-th path to a distinct offset $\mathrm{loc}_r \triangleq \left(\nu_r + 2Nc_1 \tau_r\right)_N$, yielding a sparse $H_{\mathrm{eff}}$.}  {This representation supports path estimation and detection.}  {Under suitable channel and DAFT parameter assumptions, together with the required full-rank condition on codeword differences, AFDM can achieve full path diversity over doubly dispersive channels.}

\subsection{ {Advantages of AFDM for Dynamic ISAC}}
 {The aforementioned waveform structure offers several advantages, including tunable path separation, full-diversity potential in doubly dispersive channels, OFDM hardware reuse, and tunable chirp parameters for joint sensing and data transmission within a frame.}  {Table~\ref{afdm_vs_ofdm} compares four waveforms and the following discussion examines these advantages.}

\subsubsection{ {Path Separation and Achievable Diversity}}
\begin{figure}
    \centering
    \includegraphics[width=0.65\linewidth]{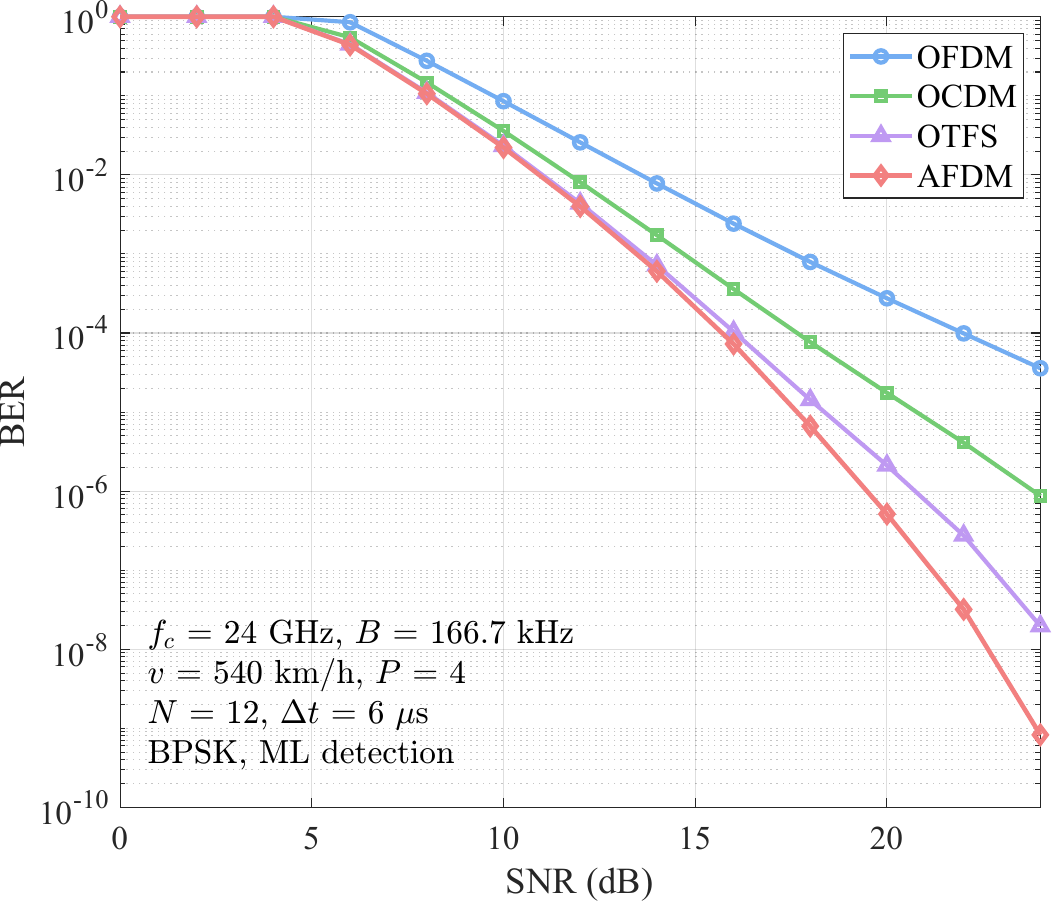}
    \caption{ {BER performance of OFDM, OCDM, OTFS and AFDM over a doubly-dispersive channel. The OTFS grid size is  \(K\times L=\text{4}\times\text{3}\) (Doppler \(\times\) delay). The AFDM chirp parameters are \(c_1=\text{3}/(\text{2}N)\) and \(c_2=\sqrt{\text{2}}/(\text{2}N)\). Each SNR point from 0 to 25 dB uses \(\text{20000}\) independent channel realizations.}}
    \label{ber}                    
\end{figure}

 {Each path's delay and Doppler determine its DAFT offset and \(c_1\) controls the delay-dependent component. Under an integer delay-Doppler model, choosing \(c_1\) to match the expected delay-Doppler support maps paths with distinct delay-Doppler pairs onto separate diagonals of the effective channel matrix, thereby separating them at the receiver, as illustrated in Fig.~\ref{yuanlitu}(c). This resolves paths that overlap under OCDM's fixed chirp slope~\cite{chirpreview}. For \(P\) independent fading paths, when the delay-Doppler pairs are distinct and the chirp parameters are chosen such that the codeword-difference matrix has rank \(P\) for every pair of distinct transmitted vectors, AFDM achieves full path diversity of order \(P\) under maximum-likelihood (ML) detection. Fig.~\ref{ber} compares BER of OFDM, OTFS, OCDM, and AFDM under ML detection in a four-path channel. At the high end of the simulated signal-to-noise ratio (SNR) range, AFDM exhibits the steepest BER decline among the four waveforms, consistent with the diversity order predicted for this configuration.}

\subsubsection{Hardware-Friendly Implementation and Compatibility}
 {A practical advantage of AFDM is its compatibility with the transform hardware already used in OFDM transceivers.}  {At the transmitter, the inverse DAFT applies chirp phase rotations before and after the IFFT.}  {The receiver applies the corresponding rotations around its FFT~\cite{afdmcore}.}  {This structure allows a dual mode baseband platform to share the same transform engines across both waveforms.}  {Chirp rotations can be enabled for AFDM and bypassed for OFDM, retaining the existing signal path.}  {Integrating AFDM also requires support for the chirp-periodic prefix and for channel estimation and equalization in the DAFT domain.}  {Furthermore, existing studies have demonstrated that AFDM systems can achieve full diversity under modeled hardware impairments~\cite{hardware}.}  {For deployment, the principal benefit is that these waveform-specific functions can be added around familiar OFDM processing while preserving core transform hardware.}

\subsubsection{ {Tunable Chirp Parameters for Sensing and Data Transmission}}
 {AFDM's tunable chirp parameters let the transmitter combine sensing and data transmission in a frame.}  {In pilot-based sensing, the receiver can estimate path delays and Doppler shifts from the response to a known DAFT-domain pilot \cite{afdmsense}.}  {The chirp slope and pilot guard region can be selected according to the expected delay and Doppler ranges.}  {A wider guard accommodates a larger spread of pilot responses at the cost of fewer resources for payload data.}  {A second use is chirp-parameter index modulation, in which the choice of a parameter configuration conveys additional information \cite{RadarCentric}.}  {The number of distinguishable configurations determines the index bits, while ambiguity sidelobes affect detection of a weak target near a stronger one. This property makes AFDM a promising candidate for ISAC.}

 {However, these benefits make AFDM sensitive to the choice of its chirp parameters.}  {If the actual delay or Doppler range exceeds the assumed range, paths may overlap in the DAFT domain and become harder to resolve.}  {The choice of \(c_2\) also affects the phase structure of the effective channel, so a configuration that separates the paths may still fail to guarantee full diversity~\cite{afdmcore}.}

\begin{table*}[t]
\centering
\caption{ {Communication, sensing, and implementation characteristics of candidate waveforms for dynamic ISAC}}
\label{afdm_vs_ofdm}
\color{blue}\small
\setlength{\tabcolsep}{4pt}
\renewcommand{\arraystretch}{1.15}
\begin{tabularx}{\textwidth}{@{}>{\raggedright\arraybackslash}p{0.11\textwidth}>{\raggedright\arraybackslash}p{0.18\textwidth}*{4}{>{\raggedright\arraybackslash}X}@{}}
\toprule
\textbf{Category} & \textbf{Metric} & \textbf{OFDM} & \textbf{OTFS} & \textbf{OCDM} & \textbf{AFDM} \\
\midrule
\multirow{2}{=}{Communication}
& Path separation~\cite{afdmcore}
& Across subcarriers and symbols
& 2D delay-Doppler resolution
& Fixed slope; possible path overlap
& Tunable slope and path offsets \\
\cmidrule(l){2-6}
& Diversity order~\cite{yin2026}
& low & comparatively high & Channel-dependent &full diversity \\
\midrule
\multirow{2}{=}{Sensing}
& Control of unambiguous range and velocity~\cite{yin2026}
& Subcarrier spacing and symbol timing
& Grid dimensions and pulse shape
& Block duration and bandwidth
& Block duration, bandwidth, and $c_1,c_2$ \\
\cmidrule(l){2-6}
& Pilot-based sensing overhead~\cite{afdmsense}
& Pilot-pattern dependent
& Pilot and 2D guard region
& Training-dependent
& Embedded pilot and guard region \\
\midrule
\multirow{2}{=}{Implementation}
& Reuse of existing OFDM baseband hardware~\cite{chirpreview}
& Native & Implementation-dependent & High & High \\
\cmidrule(l){2-6}
& Phase-noise resilience~\cite{hardware}
& Baseline
& Higher
& Higher in selected configurations
& Higher \\
\bottomrule
\end{tabularx}

\end{table*}

\section{Representative Applications of AFDM-Enabled ISAC in Dynamic Environments}
\begin{figure}
    \centering
    \includegraphics[width=0.9\linewidth]{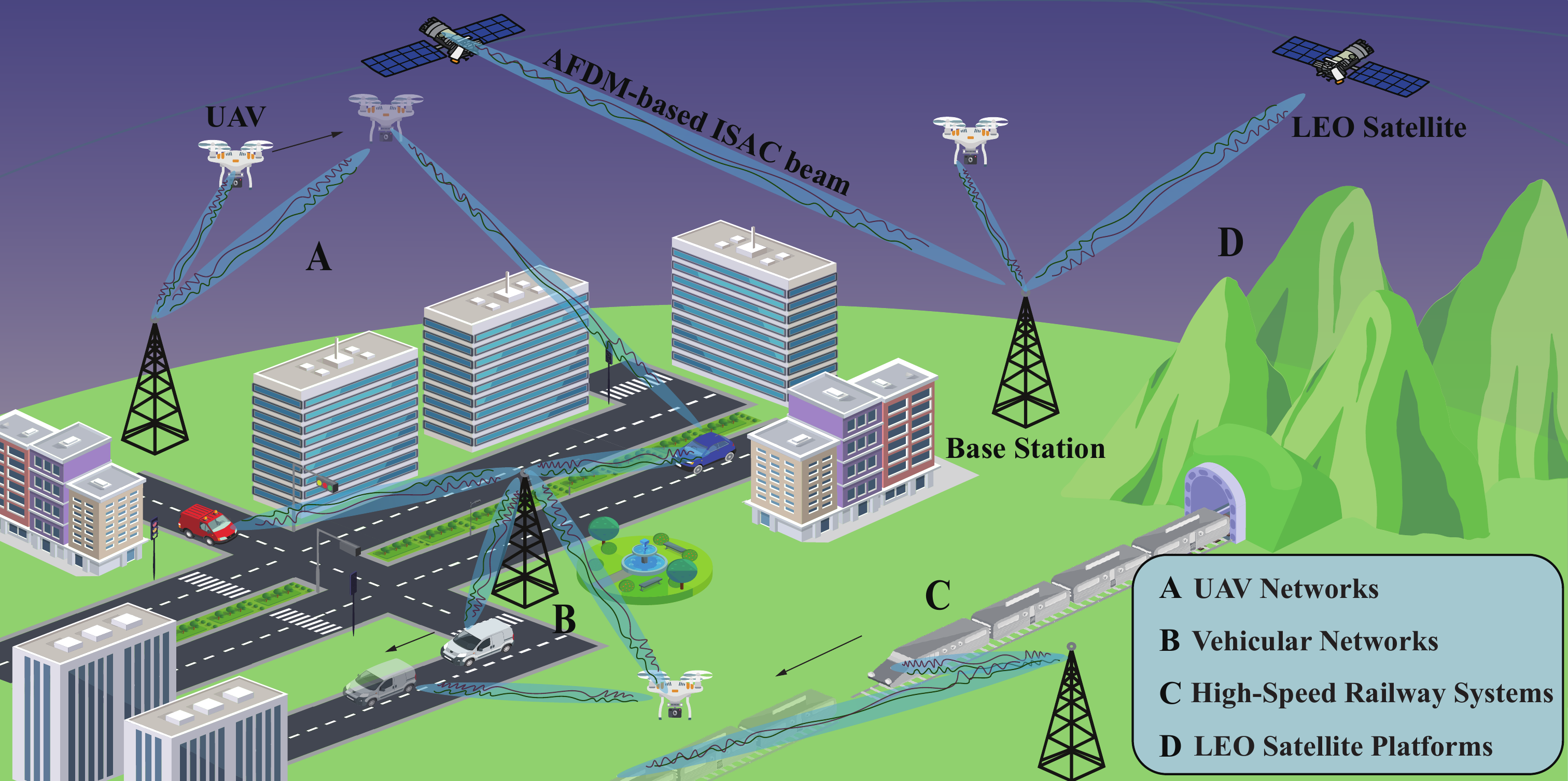}
    \caption{Representative applications of AFDM-enabled ISAC in dynamic environments.}
    \label{changjingtu}
\end{figure}
This section examines representative dynamic environments, identifies their key system challenges, and discusses the potential of AFDM to address these challenges through its distinctive delay-Doppler domain signal representation.  {As illustrated in Fig.~\ref{changjingtu}, four representative scenarios are considered: unmanned aerial vehicle (UAV) networks, vehicular networks, high-speed railway systems, and low Earth orbit (LEO) satellite platforms.}
\subsection{UAV Networks}
 {The characteristics of UAV networks include high Doppler uncertainty and arbitrarily varying velocity vectors in three-dimensional space, while the system is also constrained by the limited onboard computational resources of the platform.}  {Under these conditions, OFDM relies on the time invariance approximation within a block and suffers from severe ICI when the frequency offset is excessively large.}  {The chirp rate of OCDM is fixed at a single design point and cannot be adjusted according to the flight state.}  {Although OTFS exhibits sparsity in the delay-Doppler domain, its two-dimensional equalization imposes high demands on onboard computational resources.}  {Three features of AFDM address these constraints.}  {First, its tunable chirp parameters can be configured for the expected delay-Doppler range, enabling the receiver to distinguish and exploit propagation paths, thereby covering the Doppler variations within the flight envelope without requiring frame-by-frame redesign.}  {Second, dedicated chirp pilots can support analog de-chirping and sub-Nyquist sampling in the sensing branch, compressing wideband signals into narrowband tones and thus significantly alleviating the sampling and front-end burden of this branch.}  {Finally, the structured channel in the DAFT domain facilitates sparse channel estimation and detection, allowing the receiver to perform equalization and detection with low-complexity algorithms.}  {This translates waveform gains into an achievable onboard computational budget, supporting low-power and low-latency joint sensing and communication on resource-constrained platforms.}

\subsection{Vehicular Networks}
 {In vehicular networks, vehicle motion is constrained by road layouts and driving conditions; therefore, the expected Doppler range can be estimated by combining vehicle speed, heading direction, and road information.}  {Meanwhile, roadside buildings, vehicles, and other ground scatterers generate propagation paths with distinct delays and Doppler shifts, causing the vehicular channel to exhibit complex time-varying multipath characteristics.}  {For safety services such as collision warning, the communication system must simultaneously satisfy low latency and high reliability under short-packet transmission conditions.}  {To address these characteristics, AFDM can configure the chirp parameters according to the expected delay-Doppler range, such that different paths form structured responses in the DAFT domain that are exploitable by the receiver.}  {Vehicle status and road information can provide priors for this parameter configuration, thereby aligning the waveform design with the expected propagation environment.}  {Furthermore, the receiver can exploit the sparse structure in the DAFT domain for channel estimation and joint detection, leveraging the diversity inherent in multipath propagation.}  {For short-packet services, reducing the pilot and guard interval overhead required for channel acquisition is particularly important, as these resources consume the limited data transmission space.}  {In this context, the path structure of AFDM enables the design of low-overhead channel estimation schemes.}

\subsection{High-Speed Railway Systems} 
 {High-speed railway communication and sensing face significant Doppler effects induced by the high-speed motion of trains, as well as time-varying multipath caused by variations in the propagation environment along the railway.}  {Compared with vehicles and UAVs, which possess more flexible moving directions, trains travel along fixed tracks, and their positions, speeds, and pre-planned routes can provide prior information for Doppler range estimation.}  {Meanwhile, scenarios such as cuttings, viaducts, and tunnels exhibit distinct scattering structures, and propagation paths may strengthen, vanish, or emerge when the train passes through the boundaries between these scenarios.}  {To address these characteristics, AFDM can configure its chirp parameters by incorporating the train kinematic information and the channel information along the railway, such that the paths within the expected delay-Doppler range form distinguishable structures in the DAFT domain for the receiver to perform channel estimation and detection.}  {When multiple resolvable paths exist, AFDM can also exploit path diversity through appropriate parameter design, providing support for reliable communication under high-speed conditions.}  {For trackside target sensing, the chirp parameters can also be selected based on the expected range and velocity bounds to accommodate the sensing requirements of different railway sections.}  {As the train enters a new propagation scenario, the system still needs to update the channel information and strike a balance among reconfiguration, pilot overhead, and detection complexity according to the parameter variations.}

\subsection{LEO Satellite Platforms}
 {LEO satellite platforms face large Doppler shifts induced by the high-speed motion of satellites, long propagation delays, and link geometry that varies with satellite transit.}  {Owing to the high predictability of satellite orbits and motion states, the system can utilize ephemeris and terminal position information to estimate and pre-compensate for the dominant Doppler components.}  {After compensation, orbit and position estimation errors may introduce residual frequency offsets, while terminal motion and ground scattering may further induce multipath Doppler spread.}  {To address these effects, AFDM can configure its chirp parameters according to the expected delay-Doppler range, enabling the receiver to exploit the structured path responses in the DAFT domain for channel estimation and detection.}  {In addition, terminals in LEO networks need to initiate access in the presence of frequency offsets.}  {AFDM chirps can be utilized for random access preamble design, and existing study~\cite{LEO} have shown that under the considered satellite link configurations, AFDM preambles can mitigate the impact of frequency offsets on access detection.}  {Furthermore, AFDM can be combined with multi-user access techniques such as sparse code multiple access, providing a design space for massive connectivity in high-speed scenarios.}

\section{Key Technologies}
\subsection{Accurate Parameter Estimation}

Accurate parameter estimation is fundamental to AFDM-enabled ISAC, as the delay and Doppler shifts of propagation paths determine both communication reliability and sensing accuracy~\cite{ParametEst}. In dynamic environments, however, rapidly varying channels, closely spaced paths, and changing propagation conditions make accurate and timely parameter extraction challenging.  {Moreover, conventional methods often rely on prior knowledge of the number of paths, which may become unreliable as the environment evolves.} AFDM provides a structured signal representation in the delay-Doppler domain, offering a natural basis for joint estimation of path delays and Doppler shifts.  {Sparse Bayesian learning has been applied to low-complexity channel estimation in vehicular AFDM links by exploiting the channel structure~\cite{VTX1}.} Meanwhile, tensor decomposition can exploit the multidimensional structure of AFDM signals to achieve accurate and computationally efficient parameter estimation~\cite{chen}. By jointly identifying the delay, Doppler, and path correspondence, these techniques can avoid explicit path matching and provide timely channel and target information for subsequent beamforming, sensing, and resource adaptation.

\subsection{Dynamic Beamforming Design }

Beamforming design is critical for AFDM-enabled ISAC systems, especially in rapidly varying environments. By jointly optimizing the amplitude and phase across antennas, transmit energy can be focused toward intended users and sensing targets while suppressing inter-user and inter-target interference. More importantly, AFDM introduces additional waveform design freedom through its configurable chirp parameters, which can be dynamically adjusted according to instantaneous channel and target characteristics. In highly dynamic scenarios, user mobility and target motion lead to rapid changes in channel states and optimal beam directions, requiring beamforming parameters to be updated in real time. By jointly adapting the spatial beamforming weights and chirp parameters, AFDM can track moving users and targets, maintain accurate beam alignment, and mitigate Doppler-induced interference.  {This dynamic design provides additional flexibility beyond conventional beamforming, enabling rapid adaptation to changing environments while balancing communication and sensing requirements.}

\subsection{Index-Modulated AFDM Transmission}

Index modulation (IM) provides an additional degree of freedom for AFDM-enabled ISAC by conveying information through the activation or selection of transmission resources. In dynamic environments, this index domain can be jointly exploited with the inherent delay-Doppler and chirp domains of AFDM to adapt the transmitted waveform to rapidly changing channel and sensing conditions. Specifically, IM-AFDM can encode index bits by selecting different chirp parameter pairs from a predefined lookup table, preserving the signal structure and favorable ambiguity characteristics for sensing~\cite{RadarCentric}. Alternatively, index information can be conveyed by selectively activating resources in different signal domains~\cite{IMcode}. The resulting index-domain flexibility can further enhance AFDM-based dynamic ISAC. For communication, index modulation enables additional information transmission without relying solely on constellation symbols, offering potential gains in spectral and energy efficiency. For sensing, dynamic index selection provides an additional mechanism to adapt the transmitted signal to changing interference, propagation, and target conditions. By jointly exploiting the index, chirp, and delay-Doppler domains, IM can provide greater waveform flexibility in AFDM and enable more adaptive communication and sensing in highly dynamic environments.


\section{ {Case Study: Joint Parameter Estimation and Predictive Beam Tracking in Vehicular Networks}}
 {We consider an AFDM-enabled ISAC system in vehicular networks, where a moving vehicle transmits uplink reference signals to a fixed roadside base station (BS). The BS estimates the propagation parameters and exploits the acquired angular information for downlink beam tracking.}  {The vehicle travels along a straight road at $\text{28}$~m/s, with a perpendicular distance of $\text{12}$~m from the BS.}  {The channel contains a line-of-sight (LoS) path and a single-reflection path through a fixed scatterer.}  {Although the vehicle maintains a constant speed, its changing geometry produces time-varying path delays and Doppler shifts, together with a continuously evolving LoS arrival angle.}  {Near the closest point of approach, the LoS angle changes most rapidly, while the LoS Doppler changes sign as the vehicle passes the BS.}  {These variations create a dynamic setting for parameter acquisition and predictive beam control.}  {Parameters are assumed constant within each short observation interval and updated across observation epochs.}  {The carrier frequency and bandwidth are 28 GHz and 20 MHz, respectively.}  {The BS employs half-wavelength uniform linear arrays with 8 receive elements and  32 transmit elements.}  {Each estimation observation comprises 10 reference blocks, each containing 12 samples, with a block interval of $\text{50}~\mu$s.}  {Parameter estimation is evaluated using snapshots sampled along the trajectory over an SNR range from -20 to 10 dB, with 200 Monte Carlo trials.}

\begin{figure}
    \centering
    \includegraphics[width=0.9\linewidth]{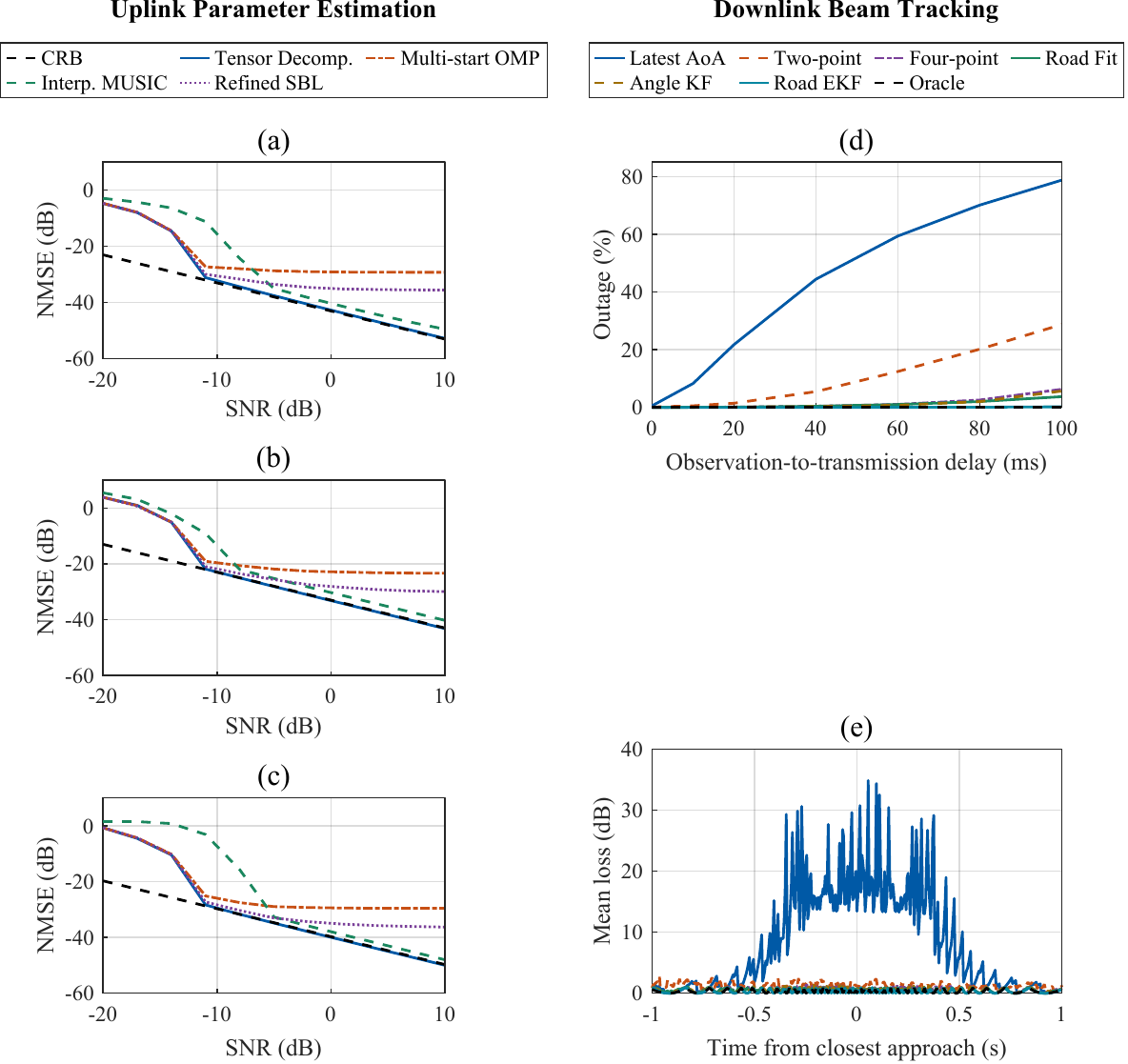}
    \caption{ {Case study of parameter acquisition and sensing-assisted beam tracking in a vehicular AFDM-enabled ISAC system: (a) normalized mean square error (NMSE) of delay estimation; (b) NMSE of Doppler estimation; (c) angle-of-arrival (AoA) estimation NMSE; (d) downlink outage probability versus observation-to-transmission delay; and (e) mean beam-pointing loss along the vehicle trajectory with a delay of 40~ms.}}
    \label{fangzhentu}
\end{figure}

 {Accurate parameter acquisition provides the foundation for environmental sensing and subsequent link control in AFDM-enabled ISAC systems.}  {Delay, Doppler, and arrival angle characterize propagation distance, radial motion, and spatial direction, respectively.}  {Fig.~\ref{fangzhentu}(a)-(c) compare their estimation NMSEs using structured tensor decomposition, multi-start orthogonal matching pursuit (OMP), interpolated multiple signal classification (MUSIC), and refined sparse Bayesian learning (SBL), together with the Cram\'er-Rao bound (CRB) for the same observation model.}  {At low SNRs, noise and path-identification errors cause substantial deviations from the CRB.}  {As the SNR increases, structured tensor decomposition approaches the CRB by exploiting the multidimensional observation structure and performing continuous parameter optimization.}  {Multi-start OMP remains competitive at low SNRs but develops an error floor as grid quantization becomes dominant.}  {Refined SBL reduces this residual error through Bayesian candidate screening, joint support selection, and limited local refinement, although its restricted refinement range leaves a remaining error floor.}  {Interpolated MUSIC exhibits a pronounced low-SNR threshold effect; once the signal subspace can be reliably resolved, its accuracy improves substantially, with interpolation reducing grid-induced errors.}

 {Beam tracking demonstrates how the acquired sensing information can support communication in a dynamic AFDM-enabled ISAC system.}  {Since vehicle motion makes historical angular information progressively outdated, the BS must predict the direction at the actual transmission time rather than simply reuse the latest estimate.}  {All practical tracking methods use the same angular sequences obtained by structured tensor decomposition.}  {As shown in Fig.~\ref{fangzhentu}(d), the outage probability of the latest-angle policy increases markedly with observation-to-transmission delay.}  {Two-point extrapolation compensates for motion but is sensitive to measurement noise, whereas four-point fitting and the angular Kalman filter improve prediction through multiple observations or recursive filtering.}  {Road-aware prediction and the road-geometry extended Kalman filter further exploit the known road geometry, providing greater delay tolerance in the considered straight-road scenario.}  {Fig.~\ref{fangzhentu}(e) shows the time-dependent mean beam-pointing loss for a fixed delay of 40~ms, revealing how the methods respond to changing viewing geometry, particularly around the closest approach.}  {The oracle reference also includes discrete-codebook quantization loss.}  {Together, these results illustrate how timely prediction and beam control translate acquired angular information into improved downlink reliability.}

\section{Open Challenges and Future Directions}

\subsection{Near-Field Coupling in AFDM-Enabled ISAC}

 {With the rapid development of extremely large-scale multiple-input multiple-output systems and high-frequency communications, an increasing number of wireless systems operate in the near-field radiation region, where spherical wavefronts and spatial nonstationarity fundamentally differ from conventional far-field channel models.} Especially in dynamic environments, the signal model exhibits increased complexity. Consequently, the delay-Doppler representation of AFDM can no longer be directly characterized by conventional planar-wave models, and the coupling among delay, Doppler, angle, and distance becomes an important issue. Extending AFDM-enabled ISAC from far-field to near-field environments therefore requires new channel modeling, parameter estimation, and waveform design frameworks.  {In particular, how to exploit AFDM's delay-Doppler-domain structure while jointly resolving range, angle, and Doppler in spherical-wave propagation remains largely unexplored.} Future research could investigate near-field AFDM waveform design, joint beam focusing and parameter estimation, and adaptive chirp parameter selection for dynamically varying propagation geometries. Such developments would provide a foundation for extending AFDM-enabled ISAC to emerging systems with large apertures and high carrier frequencies.


\subsection{Secure Transmission Strategy}

Security is an important yet relatively underexplored issue in AFDM-enabled ISAC, particularly in highly dynamic environments. The configurable chirp parameters of AFDM provide an additional waveform dimension that can be exploited for physical-layer security. For example, chirp parameter selection or hopping can be jointly designed with index modulation to conceal waveform characteristics and hinder unauthorized signal demodulation. Unlike conventional encryption mechanisms operating mainly at higher layers, such waveform-level security can be tightly integrated with the physical-layer transmission process. However, dynamically varying channels and sensing requirements introduce new challenges for secure AFDM-enabled ISAC. Future research should investigate adaptive chirp parameter design, secure beamforming, and waveform optimization against eavesdroppers while jointly considering communication rate, sensing accuracy, and energy consumption. In particular, how to dynamically balance security and ISAC performance under rapidly changing user, target, and eavesdropper locations remains an open problem. Addressing these issues could establish more robust and adaptive physical-layer security mechanisms for AFDM-enabled ISAC in dynamic environments.

%

\subsection{Hardware Impairments and Practical Implementation}

 {The practical deployment of AFDM still entails several issues that merit further investigation.}  {Although existing studies have demonstrated that AFDM can achieve full diversity under specific hardware impairment models and corresponding conditions~\cite{hardware}, this result does not imply that hardware non-idealities have no impact on its practical communication and sensing performance.}  {The preservation of diversity gain, along with the performance loss at finite SNRs, the accuracy of sensing parameter estimation, and the implementation overhead, all warrant further investigation.}  {Therefore, building upon existing theoretical results, future research should further focus on the specific implementation issues of AFDM-enabled ISAC under hardware impairments.}  {For instance, regarding phase noise, carrier frequency offset, and sampling error, it is essential to investigate the coupling relationship between synchronization errors and motion parameters, and to design robust joint synchronization, parameter estimation, and compensation methods.}  {To address the nonlinearity of radio-frequency power amplifiers and the peak-to-average power ratio  issue, signal processing schemes that jointly consider communication reliability, sensing accuracy, and computational complexity should be explored.}  {Furthermore, the resource overhead introduced by hardware compensation and reference signal configuration, as well as the impact of this overhead on the sensing update rate and real-time beam control, can also be investigated.}  {Validating the relevant methods through hardware prototypes and over-the-air experiments will help clarify the applicability conditions of existing theoretical conclusions in practical systems, thereby promoting the deployment of AFDM-enabled ISAC in real dynamic environments.}

\section{Conclusion}
 {This article has provided a systematic overview of AFDM-enabled ISAC in dynamic environments.}  {We first discussed the challenges posed by highly dynamic propagation conditions and examined the principles of four representative waveforms, with particular emphasis placed on the performance and advantages of AFDM.}  {Subsequently, representative application scenarios are presented, where their respective characteristics and the corresponding advantages of AFDM are analyzed.}  {In addition, a case study is conducted to investigate the application of these key technologies in a typical vehicular network scenario.} Finally, several open challenges were highlighted, including near-field coupling, secure transmission, and hardware impairments. Overall, AFDM provides a promising waveform-level foundation for reliable, adaptive, and efficient ISAC in dynamic environments.

\balance

\bibliographystyle{IEEEtran}
\bibliography{reference}

\end{document}